\documentclass{emulateapj}
\pdfoutput=1

\usepackage{amsmath,natbib,graphicx}
\usepackage{epsf}
\input{epsf}
\usepackage{epsfig}
\DeclareGraphicsExtensions{.jpg,.pdf,.png,.eps,.ps}
\graphicspath{{FIGURES/}}

\newcommand{\Tsz}{\mbox{$T_{\mbox{\tiny SZ}}$}}
\newcommand{\ukcmb}{\mbox{$\mu \mbox{K}_{\mbox{\tiny CMB}}$}}
\newcommand{\kcmb}{\mbox{$\mbox{K}_{\mbox{\tiny CMB}}$}}
\newcommand{\fsz}{\mbox{$f_{\mbox{\tiny SZ}}$}}

\newcommand{\thcore}{\mbox{$\theta_\mathrm{core}$}}
\newcommand{\thcoresq}{\mbox{$\theta^2_\mathrm{core}$}}
\newcommand{\snr}{\mbox{$\mathrm{S/N}$}}
\newcommand{\ltsima}{$\; \buildrel < \over \sim \;$}
\newcommand{\ltsim}{\lower.5ex\hbox{\ltsima}}

\begin{document}

\journalinfo{The Astrophysical Journal, 701:32-41, 2009 August 10}
\submitted{Received 2008 October 10; accepted 2009 June 8; published 2009 July 20}

\title{Galaxy clusters discovered with a Sunyaev-Zel'dovich effect survey}

\author{
Z. Staniszewski,\altaffilmark{1,2} P. A. R. Ade,\altaffilmark{3} K. A. Aird,\altaffilmark{4}
B. A. Benson,\altaffilmark{5} L. E. Bleem,\altaffilmark{6,7} J. E. Carlstrom,\altaffilmark{6,7,8,9} 
C. L. Chang,\altaffilmark{6,9} H.-M. Cho,\altaffilmark{5}  T. M. Crawford,\altaffilmark{6,8} 
A. T. Crites,\altaffilmark{6,8} T. de Haan,\altaffilmark{10} M. A. Dobbs,\altaffilmark{10}
N. W. Halverson,\altaffilmark{11} G. P. Holder,\altaffilmark{10}
W. L. Holzapfel,\altaffilmark{5} J. D. Hrubes,\altaffilmark{4}
M. Joy,\altaffilmark{12} R. Keisler,\altaffilmark{6,7}
T. M. Lanting,\altaffilmark{10} A. T. Lee,\altaffilmark{5,13} E. M. Leitch,\altaffilmark{6,8}
A. Loehr,\altaffilmark{14} M. Lueker,\altaffilmark{5}
J. J. McMahon,\altaffilmark{6,9} J. Mehl,\altaffilmark{5}
S. S. Meyer,\altaffilmark{6,7,8,9} J. J. Mohr,\altaffilmark{15}
T. E. Montroy,\altaffilmark{1,2} C.-C. Ngeow,\altaffilmark{15} 
S. Padin,\altaffilmark{6,8} T. Plagge,\altaffilmark{5}
C. Pryke,\altaffilmark{6,8,9} C. L. Reichardt,\altaffilmark{5}
J. E. Ruhl,\altaffilmark{1,2} K. K. Schaffer,\altaffilmark{6,9}
L. Shaw,\altaffilmark{10} E. Shirokoff,\altaffilmark{5} 
H. G. Spieler,\altaffilmark{13}
B. Stalder,\altaffilmark{14} A. A. Stark,\altaffilmark{14} 
K. Vanderlinde,\altaffilmark{6,8,9}
J. D. Vieira,\altaffilmark{6,7} O. Zahn,\altaffilmark{16} and A. Zenteno\altaffilmark{15}
}

\altaffiltext{1}{Physics Department,
Case Western Reserve University,
Cleveland, OH 44106}
\altaffiltext{2}{Center for Education and Research in Cosmology and Astrophysics,
Case Western Reserve University,
Cleveland, OH 44106}

\altaffiltext{3}{Department of Physics and Astronomy,
Cardiff University,
CF24 3YB, UK}
\altaffiltext{4}{University of Chicago,
5640 South Ellis Avenue, Chicago, IL 60637}
\altaffiltext{5}{Department of Physics,
University of California,
Berkeley, CA 94720}
\altaffiltext{6}{Kavli Institute for Cosmological Physics,
University of Chicago,
5640 South Ellis Avenue, Chicago, IL 60637}
\altaffiltext{7}{Department of Physics,
University of Chicago,
5640 South Ellis Avenue, Chicago, IL 60637}
\altaffiltext{8}{Department of Astronomy and Astrophysics,
University of Chicago,
5640 South Ellis Avenue, Chicago, IL 60637}
\altaffiltext{9}{Enrico Fermi Institute,
University of Chicago,
5640 South Ellis Avenue, Chicago, IL 60637}
\altaffiltext{10}{Department of Physics,
McGill University,
3600 Rue University, Montreal, Quebec H3A 2T8, Canada}
\altaffiltext{11}{Department of Astrophysical and Planetary Sciences and Department of Physics,
University of Colorado,
Boulder, CO 80309}
\altaffiltext{12}{Department of Space Science, VP62,
NASA Marshall Space Flight Center,
Huntsville, AL 35812}
\altaffiltext{13}{Physics Division,
Lawrence Berkeley National Laboratory,
Berkeley, CA 94720}
\altaffiltext{14}{Harvard-Smithsonian Center for Astrophysics,
60 Garden Street, Cambridge, MA 02138}
\altaffiltext{15}{Department of Astronomy and Department of Physics,
University of Illinois,
1002 West Green Street, Urbana, IL 61801}
\altaffiltext{16}{Berkeley Center for Cosmological Physics,
Department of Physics, University of California, and Lawrence Berkeley
National Labs, Berkeley, CA 94720}

\email{tcrawfor@kicp.uchicago.edu}

\begin{abstract}

The South Pole Telescope (SPT) is conducting a
Sunyaev-Zel'dovich (SZ) effect survey over large areas of the southern
sky, searching for massive galaxy clusters to high redshift.
In this preliminary study, we focus on a 40 $\mathrm{deg}^2$ area
targeted by the Blanco Cosmology Survey (BCS), which is centered  
roughly at right ascension $5^\mathrm{h} 30^\mathrm{m}$, declination
$-53^\circ$ (J2000).  Over two seasons of observations, this entire
region has been mapped by the SPT at 95~GHz, 150~GHz, and 225~GHz.  
We report the four most significant SPT detections of SZ clusters in
this field, three of which were previously unknown and, therefore,
represent the first galaxy clusters discovered with an SZ survey. 
The SZ clusters are detected as decrements with greater than $5\sigma$ 
significance in the high-sensitivity 150~GHz SPT map. The SZ spectrum
of these sources is confirmed by detections of decrements at the 
corresponding locations in the 95~GHz SPT map and non-detections at those 
locations in the 225~GHz SPT map.  Multiband optical images from the BCS
survey demonstrate significant concentrations of similarly colored
galaxies at the positions of the SZ detections.  Photometric redshift
estimates from the BCS data indicate that two of the clusters lie at
moderate redshift ($z \sim 0.4$) and two at high redshift ($z \gtrsim 0.8$).  
One of the SZ detections was previously identified as a galaxy cluster 
in the optical as part of the Abell supplementary southern cluster catalog and in the  X-ray 
using data from the \emph{ROSAT} All-Sky Survey (RASS).  Potential RASS
counterparts (not previously identified as clusters) are also found
for two of the new discoveries.  These first four galaxy clusters are
the most significant SZ detections from a subset of the ongoing SPT survey.
As such, they serve as a demonstration that SZ surveys, 
and the SPT in particular, can be an effective means for finding galaxy clusters.  

\end{abstract}


\bigskip\bigskip

\section{Introduction}
\label{sec:intro}

Galaxy clusters are the most massive collapsed objects in the
universe, and measurements of their abundance and evolution can 
be used to place constraints on cosmological models.  In particular, 
the abundance of clusters as a function of redshift depends on both
the growth of structure and the volume of space. 
A catalog of clusters with masses and redshifts can be used to constrain 
the matter density, 
the density fluctuation amplitude, and the 
dark energy equation of state \citep{wang98,haiman01,holder01b,molnar04,wang04,lima07}.

For a galaxy cluster catalog to be useful for constraining cosmological
models, the sample of clusters must be selected in a uniform and 
well understood manner \citep{melin05}.  In addition, cluster parameters---mass, in 
particular---must be estimable from the observed quantities.  Galaxy 
cluster catalogs selected via the Sunyaev-Zel'dovich effect are 
promising in both respects \citep{carlstrom02,majumdar04,melin05}.

The Sunyaev-Zel'dovich (SZ) effect is the inverse-Compton scattering of
cosmic microwave background (CMB) photons by hot plasma bound to
clusters of galaxies, which results in a distortion of the CMB
blackbody spectrum \citep{sunyaev70, sunyaev72}.  
The spectrum of the SZ effect has a null at $\sim
217\,$GHz, where there is no net distortion and the intensity remains
unchanged.  At frequencies below $\sim217\,$GHz, the spectral distortion 
results in a decrement in the CMB intensity in the direction of the galaxy
cluster.

As it is a distortion of the CMB spectrum, the surface brightness of
the SZ effect does not depend on cluster distance. Observations of the
SZ effect, therefore, permit the detection of clusters with a selection
function that is nearly redshift-independent \citep{birkinshaw99,
  carlstrom02}.  The amplitude of the SZ spectral distortion is
proportional to the integrated pressure 
along the line of sight through the cluster, and the relationship
between integrated SZ flux and cluster mass is expected to have
relatively low scatter \citep{motl05,nagai05,shaw08}. Therefore, an
SZ-flux-limited survey, combined with redshifts from optical
observations, has the potential to be a nearly ideal data set for
producing cosmological constraints.

Significant SZ detections of previously known clusters have been
reported in many tens of X-ray and optically selected galaxy clusters
with observations spanning from radio to submillimeter wavelengths
\citep[e.g.,][]{jones05,bonamente06,benson04,halverson08}.  However, despite the promise
of SZ surveys for cosmology, there has
been only one reported detection of the SZ effect in a field not
already suspected to harbor a cluster.  A deep survey of a relatively
small field with the VLA at 8.4~GHz produced what appeared to be a
significant flux decrement \citep{richards97}. However, subsequent
BIMA observations at 30~GHz were found to be inconsistent with the
interpretation of the VLA decrement as being due to the SZ effect \citep{holzapfel00b}.

The South Pole Telescope (SPT) is the first
instrument to demonstrate the requisite combination of resolution, mapping speed, and 
observing time to carry out a successful search for distant galaxy clusters 
through the SZ effect.
This paper presents the four most significant detections of SZ
clusters in a preliminary analysis of a $\sim40$ $\mathrm{deg}^2$ subset
of the SPT survey area.  

A brief description of the telescope and receiver
and descriptions of the SPT observations, data reduction, and cluster-searching 
algorithms are found in Section~\ref{sec:obs-reduc}.  The cluster candidates 
are presented in Section~\ref{sec:results}, including a discussion of 
possible X-ray counterparts in Section~\ref{sec:xray}
and optical followup observations in Section~\ref{sec:optical}.
Finally, we summarize the main results of this work in Section~\ref{sec:conclusions}.

\section{Instrument, Observations, and Data Reduction}
\label{sec:obs-reduc}

\subsection{Instrument}
\label{sec:inst}
The SPT is a 10 m diameter off-axis
telescope optimized for observations of fine angular scale CMB
anisotropy \citep{ruhl04,padin08,carlstrom09}.  The SPT receiver is
equipped with a 960-element array of superconducting transition edge
sensor (TES) bolometers, 
read out by a frequency-domain-multiplexed system using
superconducting  quantum interference device (SQUID) amplifiers.  
This
array is made up of six sub-arrays, each of which can be
configured to observe in one of the 95, 150, or $225\,$GHz
atmospheric windows.  For the observations used in this paper, the 
fractional bandwidths in the three observing bands were 
$\sim 30 \%$ at 95~GHz,
$\sim 25 \%$ at 150~GHz, and
$\sim 22 \%$ at 225~GHz.

The main lobes of the beams in the final maps used in this paper are reasonably approximated
by two-dimensional elliptical Gaussians with average full widths at half-maximum
(FWHM) of 1.5, 1.2, and 1.1 arcmin for 95, 150, and 225~GHz.  
These are larger than would be naively calculated from the aperture 
diameter and observing wavelength due to the combination of an 
underilluminated primary (to reduce spillover) and telescope 
pointing uncertainty.  (See Section \ref{sec:maps} and \citet{padin08}
for details.)  There is evidence in observations of very bright sources 
for a large ($\sim 20^\prime$-radius), low-amplitude ($\sim -35 \mathrm{dB}$) 
sidelobe, accounting for up to $20\%$ of the degree-scale beam response, 
but this sidelobe should have no impact on the arcmin scales important 
for the cluster detections presented here
(and indeed, including it in the analysis in this work results in 
percent-level changes in the results).

In a typical observation used in this paper, the instantaneous
per-detector sensitivity in the three bands was 
$\sim 500 \ukcmb \sqrt{\mathrm{s}}$ at 150~GHz and
$\sim 1000-1500 \ukcmb \sqrt{\mathrm{s}}$ at 95 and 225~GHz.\footnote{Throughout
this work, \kcmb \ refers to equivalent CMB fluctuation
temperature, i.e., the fluctuation on top of a 2.73K blackbody in the sky that 
would produce the equivalent signal at the detector.}

\subsection{Observations}
\label{sec:observations}
The SPT has begun an 
extensive survey of the high-galactic-latitude sky visible from the South
Pole.  
The results presented in this work are based on observations of a single field,
using data collected over two seasons.  
In the combined data set, the field is fully covered in our three
frequency bands.  
For each of the few hundred individual
observations\footnote{In this work, we refer to an observation as a
single-pass map of the entire field.}, 
initial data cuts are performed, various timestream
processing steps are applied, and a single map is made using the data
from all selected detectors in a given observing band.  The maps from
each individual observation are combined to produce a single map of the
observed region in each band.  Details of the data reduction up to the
final maps are given in Section \ref{sec:reduction}.  Finally, the
most sensitive single-frequency map is searched for SZ cluster
candidates using a matched-filter technique, the details of which are
given in Sections \ref{sec:cluster-extraction} and \ref{sec:falsedet}.

The observations included in this analysis were performed during the
2007 and 2008 SPT observing seasons, using a different focal plane
configuration each season.  Both the 2007 and 2008
focal planes included detectors sensitive to radiation within bands 
centered at approximately  95~GHz, 150~GHz, and 225~GHz. 
Details of the telescope design and performance, focal-plane
configuration, and observing strategy for these two seasons are
described in \citet{carlstrom09}.  During the 2007 Austral Winter,
the SPT began a survey of a $\sim40$ $\mathrm{deg}^2$ field that was
also targeted by the Blanco Cosmology Survey\footnote{\tt http://cosmology.uiuc.edu/BCS/} 
(BCS) project. This field is centered roughly
at right ascension (R.A.) $5^\mathrm{h} 30^\mathrm{m}$, 
declination (decl.) $-53^\circ$ (J2000)
and is hereafter referred to as the BCS5h30 field. The observations of the
BCS5h30 field with the 2007 SPT focal plane that are included in this paper
took place between 2007 July 27 and 2007 September 19, 
for a total of 280 hr of observing
time. From this period, we use only the 95~GHz data taken with 
detectors that passed a set of performance cuts,
yielding a median of 129 well-performing detectors per observation.

Between 2008 February 29 and June 5, a $\sim90$ $\mathrm{deg}^2$ 
field including the BCS5h30 field was observed with the 2008
SPT focal plane.  For this paper, 
we include 607 hr of observing time on the $\sim90$ $\mathrm{deg}^2$ 
field  with a median of 322 good 150~GHz detectors and 170 good 225~GHz
detectors.  We report results only for the subset
of this larger field that overlaps the BCS5h30 field.  

The observations reported here were performed using constant-elevation
scans across the field.  In this scan strategy, the entire telescope
is swept at constant angular velocity in azimuth from one  
edge of the field to the other, and then back.  After each such pair
of scans across the field, the telescope executes a step in
elevation before performing the next pair.  A scanning speed of 0.84 deg 
of azimuth per second 
and a scan throw of 13.5 deg of azimuth 
were chosen for the 2007 observations presented
here, 
resulting in a total scan length of 38.7 s (including 
turnarounds).  The 2008 observations employed two different scan speeds, 0.44
and 0.48 deg of azimuth per second and a scan length of 17.5 deg 
in azimuth, resulting in total scan lengths of 75-80 s (including 
turnarounds).  
The size of  
the elevation step between pairs of scans was 0.07 deg for the 2007 
observations and 0.125 deg for the 2008 observations.  
A map of the entire field
is made using this strategy, and we refer to such a single-pass map 
as a single observation.  One observation takes up to 2 hr.
The short time period for a single observation allows
for a conservative schedule of interleaved calibrations and 
facilitates data selection and reduction. Each
individual observation produces a fully sampled map of the field, 
but not fully sampled by each individual detector.  
A series of different starting elevations are
used for successive observations to provide even, fully sampled,
coverage of the field over several days.

Between individual observations of the field, we perform a series of
short calibration measurements described in more detail in 
\citet{carlstrom09}.  These include measurements of a chopped thermal source, $\sim2$ deg elevation nods,
and scans across the galactic HII regions RCW38 and MAT5a.  This
series of regular calibration measurements allows us to identify
 detectors with good performance, assess relative detector gains, monitor atmospheric
opacity and beam parameters, and constrain pointing variations.  

\subsection{Data Reduction}
\label{sec:reduction}

For this initial cluster-finding analysis, a preliminary data
reduction pipeline was developed that provides maps of the BCS5h30
field that are well 
understood, but not yet optimized.  Details of the data processing
as presented below are subject to improvement for future analysis.

\subsubsection{Data Selection and Notch Filtering}
\label{sec:cuts}

The first step in the data reduction process is to identify the data
that will be included in each single-observation map.  For every
observation, a set of well-performing detectors is identified,
primarily by assessing each detector's response to the chopped thermal
source, its response to atmospheric emission during the $\sim2$
degree elevation nods, and its noise in the frequency band appropriate  
for cluster signals.  Performance is also assessed based on
the shape of the individual detector's noise power spectrum.  If the
power spectrum has too many lines or other deviations from the
expected functional form, that detector is omitted from that
observation's analysis.  The median number of detectors stated in
Section \ref{sec:observations} is obtained after the application of these
criteria.

In addition to cutting all data from individual detectors with
anomalous noise power spectra, a small amount of bandwidth is cut from
all detectors in certain observations.
The receiver exhibits sensitivity to the pulse-tube cooler, 
resulting in occasional lines
in the detector noise power spectra 
at frequencies corresponding to the pulse-tube frequency 
(approximately 1.6~Hz) and its harmonics.  
In every observation, all detectors' noise power spectra are combined
in quadrature, and a search is performed for features in the resulting spectrum 
at every harmonic of the pulse-tube frequency.  If a high-significance
feature is found at any harmonic, a notch filter 
around that harmonic is applied to every detector's timestream.  
The width of the notch is determined by the fit, 
with a maximum of 0.007~Hz full-width. The maximum amount of 
bandwidth that could be cut with this filter is $0.4\%$, but the actual 
amount is far smaller.

The data are eventually parsed into individual azimuth scans and are
further selected for inclusion in the analysis on a scan-by-scan basis. 
Only the constant-azimuthal-velocity portion of each scan is eligible
for inclusion; data taken while the telescope is accelerating
are omitted.  Scans during
which there were data acquisition problems or large ($> 20$ arcsec) 
instantaneous
pointing errors---roughly $5 \%$ of total scans---are flagged for 
omission.  For each detector, data
from an individual scan are flagged if the detector or its associated 
readout channel exhibits high noise, 
if the demodulated detector output comes close to the limits of the 
analog-to-digital converter, 
if the readout SQUID associated with the detector exhibits an 
anomalously large DC offset, or if the detector
demonstrates any cosmic ray-like events in its time-ordered data.  
Typically about $5 \%$ of all otherwise well-performing 
bolometers are flagged within a given scan for one of these reasons.
Data that remain unflagged after all of these cuts have been applied
are processed and included in the maps.

Finally, individual observations are assessed for quality before
inclusion in the final coadded maps.  Observations are excluded
if the thermal conditions of the receiver were not 
sufficiently stable, if the number of 
well-performing detectors was anomalously low, or if the observation
was not fully completed.  Individual observation 
maps are also excluded from the final maps if 
the rms in the single-observation map exceeds an empirically 
determined threshold.  Of the complete observations of the BCS5h30 field,
approximately $77 \%$ (270/350) of the 2007 observations
and $83 \%$ (314/377) of the 2008 observations are included in the 
final maps.

\subsubsection{Relative and Absolute Calibration}
\label{sec:calibration}

The relative gains of the detectors, and their gain variations over time, are
estimated using measurements of their response to a chopped thermal source
taken before every observation.  This relative calibration is used to
``flat-field'' the array before estimation of any common-mode
timestream signal and before mapmaking.  
For the relative calibration step before mapmaking, we apply
corrections to account for variations in illumination of the thermal
source across the focal  plane.  These are measured by calculating the
ratio of each detector's response to the thermal source to its
response to the galactic HII region RCW38, averaged over many measurements.  

We convert the RCW38-corrected relative calibration into an absolute
calibration by comparing our 
measurements of RCW38 with published flux measurements in 
our observing bands.  For 150 and 225~GHz, we compare 
to the RCW38 flux reported by ACBAR \citep{runyan03a},
which was recently linked to the WMAP5 calibration at 
150~GHz \citep{reichardt09}.  In order to use the ACBAR
calibration, we smooth the SPT map of RCW38 to match the
ACBAR resolution.  We then compare the flux within the 
inner 8 arcmin of the SPT map with the ACBAR-determined flux.  For the
95~GHz calibration, we use the flux reported by BOOMERANG 
\citep{coble03}.  In this case, we smooth the SPT 
RCW38 map to match the BOOMERANG resolution and then
compare the flux within two times the Gaussian width 
($\sigma$) of the BOOMERANG beam.

This calibration method does not correct for
temporal variations in atmospheric opacity between
SPT measurements of RCW38.  This results in a slightly sub-optimal weighting 
of detectors during map-making, but does not bias the absolute calibration.  
The added absolute calibration uncertainty from ignoring atmospheric 
opacity corrections is small relative to our overall calibration uncertainty.  
For each of our bands over both seasons, the rms variation in the brightness of
RCW38 is measured to be $< 5\%$, even without correcting for differences 
in atmospheric opacity or changes in the relative response of the bolometers.

We estimate the
calibration uncertainty of the 95, 150, and 225~GHz bands to be 
$13\%$, $8\%$, and $16\%$,
respectively.  At these levels of uncertainty, the 
RCW38-based calibration is consistent with other cross-checks 
of our calibration, including initial CMB power spectrum 
estimates and observations of Neptune.

\subsubsection{Timestream Processing}
\label{sec:processing}

In addition to the notch filtering described in Section 
\ref{sec:cuts}, the timestream processing of detector data 
consists of deconvolution of the detector response function, 
low-pass filtering, and projecting out long-timescale drifts 
due to atmosphere and readout $1/f$ noise.

Detector-response deconvolution and low-pass filtering are 
done in a single step.  The detector temporal-response functions are measured 
periodically by sweeping 
the chop frequency of the thermal calibration source and measuring
the amplitude and phase of each detector's response.  These temporal-response 
functions are fit adequately by a single-pole low-pass filter, and the 
time constants do not vary significantly from observation to 
observation.  Across the focal plane, the time constants ranged
from 5 to 10 ms in 2007 and from 10 to 20 ms in 2008.
The cutoff for the applied low-pass filter is set at 40~Hz for the 2007
data and 25~Hz for the 2008 data, such that in conjunction with a digital 
filter already applied by the data acquisition 
computer, they filter approximately the same spatial scales 
($\lesssim 0.5$ arcmin), given the
different scan speeds for each season.  This combination acts as an 
anti-aliasing filter for our eventual resampling of the data onto 
0.25 arcmin map pixels but does not suppress power on scales 
of the SPT beams.

For every scan, each detector's timestream is then fit
simultaneously to a number of slowly varying template functions, and
the best fit to each template is subtracted from that detector's
timestream.  This time-domain filter reduces the effect of low-frequency
noise in the detector timestreams due to readout noise or to
atmospheric fluctuations.   For this analysis, we use a combination of Legendre
polynomials and an array common mode as the templates.  
The highest order of Legendre polynomial used is 19 for the 2008 data and 11 
for the 2007 data, resulting in a characteristic filter scale of 
roughly one-half degree for both seasons.

The common-mode template is constructed from the mean across all the
well-performing detectors in the array at a given observing frequency,
using a nominal relative calibration.  Removing this common mode
should eliminate the majority of the atmospheric fluctuation power in the detector
timestreams, because this atmospheric signal is highly correlated
between detectors.  The common-mode subtraction
acts as a spatial high-pass filter with a characteristic scale that 
roughly corresponds to the 1 deg angular size of the array.  

In order to avoid contamination of the best-fit
common mode and polynomial by bright point sources, 
sources were detected in a preliminary set of maps, 
and the locations of those sources were
masked in the common mode and polynomial subtraction algorithm.
A total of 92 sources were masked in this analysis, 35 
of which lie in the 
$\sim40$ $\mathrm{deg}^2$ BCS5h30 field.
A mask radius of 5 arcmin was used for the five brightest 
sources (including the three brightest in the BCS5h30 field),
and a mask radius of 2 arcmin was used for the rest.  

\subsubsection{Mapmaking and Pointing Corrections}
\label{sec:maps}

For every observation, a map is made for each
observing frequency using the processed data for all detectors 
in that band.  Pointing information (R.A. and decl.) is 
calculated for each detector using focal-plane offsets measured
in observations of the galactic HII regions, and boresight pointing
calculated using data from the telescope pointing readout system, with
a set of corrections described below.  Pointing coordinates are
converted to pixel number using a Sanson-Flamsteed projection 
\citep{calabretta02} with 0.25 arcmin pixels, 
and all measurements of a given pixel's brightness
are averaged using inverse-variance weighting based on the 
rms of each detector's processed and relative-gain-scaled 
time-ordered data.  

Small corrections must be applied to the pointing information
in the timestream to ensure that pointing errors are suitably small compared to
the size of the SPT beams.
The largest pointing errors of the SPT are attributed to thermal gradients across the
telescope support structure. 
These pointing errors are corrected from 20 arcsec rms to better 
than 8 arcsec rms by using an offline model 
which incorporates information from thermal and linear displacement 
sensors on the telescope structure and observations of HII regions.
The astrometry of the pointing 
model is tied to the PMN and SUMSS catalogs 
\citep{wright94,mauch03} and is accurate to 10 arcsec 
in the final maps.  The main-lobe beams in the final maps are well 
approximated by two-dimensional elliptical Gaussians with average FWHM
of 1.5, 1.2, and 1.1 arcmin 
for 95, 150, and 225~GHz.  As the beams include the effects of pointing variations and the timestream filtering, they are larger 
than expected from the diffraction of the central 8 m diameter region of primary mirror illuminated by the SPT-SZ optics \citep[see][]{padin08}.

\subsection{Identifying Galaxy Cluster Candidates}

\label{sec:cluster-extraction}

In this initial analysis, we identify the highest significance
cluster candidates using only the 150~GHz map, which has the highest sensitivity.  
The maps at the other two frequencies are used to provide a cross check of the cluster detections.  
The cluster thermal SZ signal should nearly vanish in the 225~GHz map, while the 95~GHz map can be used 
to confirm the detected clusters.  We enhance the signal to noise in the SPT maps for sources 
with morphologies similar to that expected for SZ galaxy
clusters through the application of matched spatial filters
 \citep{haehnelt96,herranz02a,herranz02b,melin06}. The matched
 filter, which we apply in the Fourier domain, combines knowledge of the source template with a noise
 estimate to optimize the signal-to-noise of the source in the
 filtered map: 
 \begin{equation}
   \psi = \frac{S^T N^{-1}}{\sqrt{S^T N^{-1}S}}
  \end{equation}
where $\psi$ is the matched filter, $N$ is the noise covariance matrix
(including non-SZ foregrounds) and $S$ is the assumed source
template. 

 \subsubsection{Source Templates}
The optimal choice for the source
template is unclear for SZ clusters, so we explore three basic
profiles, all motivated by some guess at the underlying 
three-dimensional pressure profile of clusters: 
a modified version of the Navarro-Frenk-White \citep[NFW;][]{navarro96,navarro97}
model as suggested in \citet{nagai07}, numerically integrated 
along the line of sight to 
produce a model for the SZ surface brightness; 
a Gaussian profile; 
and a $\beta$-model with $\beta$ between $2/3$ and $4/3$, which 
can be analytically integrated to yield the SZ surface brightness. The three
models proved almost indistinguishable in terms of detecting the
highly significant clusters presented in this work.  The results in
Section \ref{sec:results} are based on a spherical $\beta$-model with
$\beta = 1$.  
The SZ surface brightness profile using a $\beta$-model is given by
\begin{equation}
\Delta \Tsz (\theta) = \Delta T_{0}(1 + \theta^2/\thcoresq)^{(1-3\beta)/2},
\end{equation}
where $\theta$ is the angular distance from the line of sight through 
the center of the cluster,
$\thcore$ is the (angular) core radius, and $\Delta T_{0}$ is the peak signal.
 We choose to fix $\beta$ and vary \thcore, as these two parameters
are highly degenerate in the fit, given the current depth and
resolution of our observations.   The model is
truncated at $10 \times \thcore$.  We have explored the dependence on the scale
size, parametrized as \thcore \ in the range from
$0.25^\prime$ to $3.5^\prime$ in $0.25^\prime$ steps, and generated a matched filter of each case. 

The timestream filtering described in Section \ref{sec:processing}
affects the expected shape of the cluster signal in our maps.  To
account for this effect, we convolve the source templates with a spatial
filter intended to represent the map-domain version of our timestream
filtering before including the source templates in the matched filter
for cluster detection.  The spatial filter used in this work is simply
a high-pass in the R.A.~direction, with a cutoff on roughly half-degree
scales.  Simulations of the timestream filtering have shown this to be
an accurate representation of the effect of our timestream polynomial
removal.  We do not attempt to include the effects of our common-mode
subtraction, which predominantly affects modes that will be heavily
de-weighted in the matched filter by the inclusion of the primary CMB
as a noise term.

 \subsubsection{Noise and Foreground Estimates}
 \label{sec:noise}
 In addition to the SZ signal from massive clusters, SPT maps contain
 primary CMB, instrument and atmospheric noise, 
 point sources, and an unresolved SZ background.  For cluster detection, 
 all of these are noise sources. 
The covariance matrix used to create the optimal
 matched filter can be expressed as the sum, $N =
 N_\mathrm{CMB}+N_\mathrm{noise}+N_\mathrm{PS}+N_\mathrm{SZ}$, of these components. 
The point-source covariance only includes the unresolved 
source background; 
 point sources bright enough to be detected in our maps are treated 
separately as discussed below.
 
  The signal covariance of the primary CMB and undetected point sources
  is estimated analytically from the best-fit WMAP5 CMB power spectrum
  \citep{nolta09} and the \cite{borys03} model for dusty point sources. 
We assume that the contribution of faint radio sources to the background
source covariance is negligible. 
We assume a
spectral index of 2.7 to extend the dusty source counts to 150~GHz.  The
power spectrum of this source count distribution is estimated with the
formalism in \cite{white04}.  
The SZ background covariance is estimated 
from the simulations described in \citet{shaw08}.  

 A different approach is taken for the bright  point sources in the map. 
 Any bright sources left in our map will cause ringing in the
 maps due to the matched filter used, which can give spurious high
 significance decrements in their 
 vicinity.  We take advantage of the fact that these sources show up
 as bright spots in our maps, while any SZ cluster will be a decrement 
 at 150~GHz. We filter the 150~GHz map using a simple matched filter
 assuming isotropic noise and no signal filtering, and we flag every 
 positive source above $5\sigma$ in the filtered map, totaling 47 sources 
 in the $\sim40$ $\mathrm{deg}^2$ BCS5h30 field.  (Note that this number 
 differs from the number of sources detected in a preliminary map and
 masked in the timestream processing.  The mapmaking, source flagging,
 and timestream processing loop was not iterated over.)  All pixels within 
 a radius of four times the beam FWHM of the locations of these sources
 are set to a DC level  
 defined to be the average of the pixels just outside the mask radius.
 To avoid any edge effects from the masking procedure, pixels within
 8$^\prime$ of a masked source 
  are flagged and excluded from later cluster finding.  
 This technique proved adequate to avoid false
 detections of decrements near bright positive sources.

 The instrumental and atmospheric noise properties of the maps are
 estimated using the two-dimensional power spectrum of jackknife noise
 maps \citep{sayers09,halverson08}.  Under the assumption of
 stationarity in the map basis, the noise covariance matrix
 $\tilde{N}_{noise}$ is diagonal in the Fourier domain and equal to
 the noise power spectrum.  
We construct an individual jackknife map by
multiplying one half of the individual observations
(selected randomly) by $-1$ then coadding the full set of observations.
We repeat this process $n > m$ times (where $m$ is the number of individual
observations), computing the 2d spatial
 power spectrum for each individual jackknife map.  The mean of these
 power spectra is our estimate of $\tilde{N}_{noise}$.  

\subsubsection{Significance}
\label{sec:Significance}

  The significance of a detection is estimated using the standard
  deviation of the filtered map pixels while excluding regions around
  bright point sources.  As the integration time per pixel is a weak function
  of elevation, the scatter is estimated independently for each
  1.5$^\circ$ band in elevation. The coverage of the selected region
  is uniform in azimuth. We define detection significance as the ratio of the
  peak of the candidate decrement divided by the rms at the appropriate
  elevation.

\subsection{False Detection Simulations}
\label{sec:falsedet}
To assess the likelihood of false detections in our deep
150~GHz maps, we run the cluster detection pipeline over a series
of simulated maps.  The simulated maps include models of astrophysical
contaminants (primary CMB and point sources) and noise.  The CMB
contribution to the simulated maps is a realization of the WMAP5 CMB
power spectrum \citep{nolta09}. The point-source contribution is a
spatially random distribution of sources drawn from models based on
the \citet{borys03} counts for dusty protogalaxies and a 
version of the \citet{toffolatti98} counts for radio-loud active galactic 
nuclei (AGN) modified 
to match the DASI, CBI, and VSA counts at 30~GHz \citep{kovac02,mason03,cleary05}.  
(As predicted in Section \ref{sec:noise}, the contribution of the unresolved
AGN population to the power in the simulated maps turns out to be 
negligible.)
Both
the CMB and point-source contributions are spatially filtered to mimic
the effect of our beams and most of our timestream processing. The
effect of the common-mode subtraction is not included.  The noise in
the simulated maps is created using a jackknife procedure identical 
to the one described in Section \ref{sec:noise}.  
As with the real data, clusters are searched for in each of the 14 
filtered versions of each simulated map, using a $\beta$-model 
source template with $\beta=1$ and \thcore \ 
ranging from $0.25^\prime$ to $3.5^\prime$ in $0.25^\prime$ steps.
In 100 simulated maps, we incur an average of 0.02 false
detections per $40 \mathrm{deg}^2$ for the signal-to-noise ratio of the least significant 
cluster candidate listed in
Table \ref{tab:clusterlist} and Figure \ref{fig:cluster_plot}.  
Furthermore, a disproportionately large fraction of the false 
detections occur when the simulated maps are filtered with the 
smallest-scale optimal filter --- which is not surprising, given that
there are more independent resolution elements in this filtered 
map than in the maps filtered for detection of larger-scale objects.
If we restrict the filtering of the simulated maps to the scales on 
which the four clusters in Table \ref{tab:clusterlist} are most 
effectively detected, the false rate above $5.5 \sigma$ drops 
below 0.01 per $40 \mathrm{deg}^2$.

\section{Results}
\label{sec:results}
In this section, we present the four highest significance cluster candidates found by
our matched filter in the 150~GHz map of the $\sim40$ $\mathrm{deg}^2$ BCS5h30 field.  
The highest significance candidate in
our field was previously identified as a cluster in the Abell supplementary southern
catalog \citep{abell89}, in which it is identified as AS0520, 
and in the ROSAT-ESO Flux Limited X-ray (REFLEX) survey 
\citep{boehringer04}, in which it is identified as RXCJ0516.6-5430.  
The remaining three cluster candidates are new discoveries.  
Based on the simulated observations described in Section \ref{sec:falsedet}, 
there is roughly a $2\%$ chance that our lowest-significance 
detection is false; the chances that any of the top three are false 
detections are significantly smaller than $1\%$.

To confirm that these detections have a thermal SZ spectrum, we
look in filtered versions of our 95~GHz and 225~GHz maps at the
locations of the 150~GHz detections.  To check each object at the other two
frequencies, we use a single-frequency filtered map --- created using a matched
filter as described in Section \ref{sec:cluster-extraction} with the
cluster spatial profile that produced the highest significance for 
that object in the 150~GHz map --- and look at the single map 
pixel that corresponds to the center of the 150~GHz detection.  The use of this 
method, and the fact that the pixel distributions in the filtered maps 
are Gaussian to a high degree of precision, allows us to easily interpret 
the signal-to-noise ratio (S/N) at the cluster candidate locations 
in the 95~GHz and 225~GHz maps.

The candidate locations, the S/N at that location in the
filtered map in each of our observing bands, 
the value of \thcore \ that maximized the S/N at 150~GHz,
and the value of the best-fit central Comptonization parameter, $y_0$,
for that value of \thcore \ 
are presented in Table
\ref{tab:clusterlist}.  
We emphasize that this value of \thcore \ is the result of a search 
over a very coarse grid in parameter space using a model that may not be an accurate 
description of the detailed cluster morphology.  It is also reported
without a confidence interval.  As such, it should be interpreted only
as a rough  measure of the angular scale of the cluster candidates.
Similarly, the value of $y_0$ and the uncertainty on that value 
reported in the table are the results of an effective one-parameter fit
to the data with $\beta$ fixed at 1 and \thcore \ fixed at the grid 
value that maximized detection significance.

\begin{figure}
\begin{center}
\includegraphics[width=0.5\textwidth]{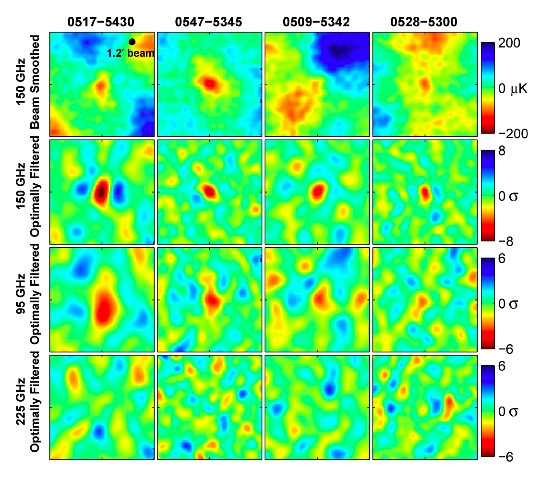}
\caption{Images of four galaxy clusters found in the SPT SZ survey.  In each panel, the region
shown is a 20 by 20 arcmin box centered on the cluster.  
All images are
oriented with north up and east to the left.
In the top row, we show the beam-smoothed 150~GHz map, and the scale has units of \ukcmb.  The lower three rows show the 150, 95 and 225~GHz maps filtered
using a $\beta=1$ model with the \thcore \ listed in Table \ref{tab:clusterlist}.  The ringing on either side of the cluster in the 150~GHz filtered maps is an artifact of the filtering.  The scale gives detection significance in $\sigma$.  
The detections
at 95~GHz range in significance from $2.5 \sigma$ to  $3.9 \sigma$
(see Table \ref{tab:clusterlist}) and provide supporting evidence for the 150~GHz
cluster detections.  
Our 225~GHz
maps are consistent with noise at these four locations, 
(see Table \ref{tab:clusterlist} and Section
\ref{sec:results}), providing another cross
check that the data are consistent with SZ sources. 
The first cluster shown here, SPT-CL 0517-5430, was previously identified
in the REFLEX X-ray cluster survey, in which it is identified as RXCJ0516.6-5430
and in the Abell supplementary southern catalog, in which it is identified as AS0520.
\label{fig:cluster_plot}}
\end{center}
\end{figure}

Images of all four candidates in 
the filtered maps are shown in Figure \ref{fig:cluster_plot}.
Two things are immediately evident from the S/N values in the table and the images in
the figure: 1) From the significance of the 150~GHz detections alone, all of the candidates are
inconsistent with noise fluctuations.  
(They are also inconsistent with emission from sources such as
radio-loud AGN or dusty protogalaxies because of the polarity of the
signal.)  2) From the 95 and 225~GHz images and detection significances,
all four detections appear consistent with thermal SZ signal and
inconsistent with primary CMB, since each candidate is significantly detected 
in the 95~GHz map but not in the 225~GHz map.

We can make this second statement more quantitative.  Using the
relative depths of our 150~GHz and 95~GHz maps and the frequency dependence of the
thermal SZ effect,
we can make a prediction for the detection significance at 95~GHz from
the 150~GHz detections and for the expected rms deviation (due to noise 
and calibration uncertainty) between our predicted and observed values. 
Taking into account the absolute calibration uncertainty in each band 
(as estimated in Section \ref{sec:calibration}) and assuming the calibration 
errors are uncorrelated between bands, the expected signal at 95~GHz
given our 150~GHz measurement is:
\begin{equation}
T_{95} (\mathrm{pred.}) = T_{150} 
\frac{\fsz(95)}{\fsz (150)},
\end{equation}
with an uncertainty of:
\begin{equation}
\sigma(T_{95},\mathrm{pred.}) = \sqrt{\sigma^2_{150} + 
(\delta_{\mathrm{cal},150} T_{150})^2}
\frac{\fsz(95)}{\fsz (150)},
\end{equation}
where $\sigma^2_{150}$ is the noise in the optimally filtered 150~GHz map, 
$\delta_{\mathrm{cal},150}$ is the calibration uncertainty at 150~GHz, and  
$\fsz(\nu)$ is the value of the frequency-dependent thermal SZ brightness
relative to the CMB background at frequency $\nu$ (in GHz).  Similarly, the uncertainty 
in the observed 95~GHz signal is 
\begin{equation}
\sigma(T_{95}, \mathrm{obs.}) = \sqrt{\sigma^2_{95} + (\delta_{\mathrm{cal},95} T_{95})^2}.
\end{equation}
The expected deviation between the predicted and observed values of the 95~GHz 
S/N for a single cluster candidate will be the quadrature sum of the above uncertainty terms, 
divided by the 95~GHz noise:
\begin{eqnarray}
\label{eqn:var95}
\left \langle \left ( \Delta (\snr)_{95} \right )^2 \right \rangle &=& 
1 + (\delta_{\mathrm{cal},95} (\snr)_{95})^2 +
\left ( \frac{\fsz(95)}{\fsz(150)} \frac{\sigma_{150}}{\sigma_{95}} \right )^2 \\
\nonumber && \times \left [1 + (\delta_{\mathrm{cal},150} (\snr)_{150})^2 \right ].
\end{eqnarray}

The rms
is approximately 4.8 times higher in CMB fluctuation temperature units
in our 95~GHz map than in our 150~GHz map, but in those same units 
the thermal SZ signal should be approximately 1.6 
times stronger at 95~GHz.  Using these values, the predicted 95~GHz S/N values for our four
candidates are -3.0, -2.5, -2.0, and -1.9, as compared to the observed
values of -3.4, -3.9, -3.4, and -2.6 (compare Table \ref{tab:clusterlist}).
Using Eqn.~\ref{eqn:var95}, the rms deviation between the predicted and observed
95~GHz S/N should be 1.2, 1.2, 1.2, and 1.1 for the four candidates.
Taking into account the correlated nature of the calibration contribution in 
Eqn.~\ref{eqn:var95}, a $\chi^2$ test for the 
difference between the predicted and observed 95~GHz S/N values
produces a
probability to exceed (PTE) of 0.66.  

Because our 225~GHz band is 
near the thermal SZ null ($\fsz(225) \simeq 0.1$), 
we should see very little signal at
cluster locations in our 225~GHz map,
and the 225~GHz equivalents to the last two terms in Eqn.~\ref{eqn:var95}
should vanish.  For this reason, and because the observed 225~GHz S/N 
values are quite small, the expected rms deviation between predicted 
and observed 225~GHz S/N should be very close to unity, and the deviations
should be almost uncorrelated between cluster candidates.  A $\chi^2$ test for the
observed S/N at 225~GHz produces a PTE of 0.30.

We can do a similar
exercise for the hypothesis that the observed signal has a 2.7~K
thermal spectrum (as would be the case for primary
CMB fluctuations) by replacing the \fsz \ ratio in 
Eqn.~\ref{eqn:var95} with unity (because the maps are calibrated to 
be in units of equivalent CMB fluctuation temperature).  
In this case, the PTE for 95~GHz is still 
non-negligible ($\sim0.09$), because the difference between the spectra of thermal SZ and
primary CMB is not so great between 95 and 150~GHz. 
However, given the depth of the 225~GHz map (the rms in CMB fluctuation 
temperature is approximately 2.5 times that in the 150~GHz map), 
we would expect S/N values of -3.5, -3.0, -2.4, and -2.2 at 225~GHz
for a 2.7~K thermal spectrum.  The PTE for this model given the observed
S/N values at 225~GHz is $\sim 6 \times 10^{-6}$, giving strong evidence 
that the signal spectrum of our cluster candidates is inconsistent with primary
CMB but consistent with thermal SZ.  

\begin{table*}
\begin{center}
\caption{Cluster Detections} \small
\begin{tabular}{cccccccc}
\hline\hline
\rule[-2mm]{0mm}{6mm}
ID & R.A. & Decl. & S/N at: &  &  & Best \thcore & Best $y_0 \times 10^4$ \\
 & & & 150 GHz & 95 GHz & 225 GHz & &  \\
\hline
SPT-CL 0517-5430 & 79.144 & -54.506 & -8.8 & -3.4 & -0.4 & $1.5^\prime$ & $ 0.97 \pm 0.13$ \\ 
SPT-CL 0547-5345 & 86.650 & -53.756 & -7.4 & -3.9 & 1.9 & $0.5^\prime$ & $ 1.31 \pm 0.21$ \\ 
SPT-CL 0509-5342 & 77.333 & -53.702 & -6.0 & -3.4 & 0.1 & $1.25^\prime$ & $ 0.67 \pm 0.12$ \\ 
SPT-CL 0528-5300 & 82.011 & -52.998 & -5.6 & -2.6 & -1.3 & $0.5^\prime$ & $ 1.00 \pm 0.19$ \\ 
\hline
\end{tabular}
\label{tab:clusterlist}
\tablecomments{
R.A.~and decl.~are in units of deg (J2000).
The value of \thcore \ reported here is that which maximized the S/N 
of each cluster in the filtered 150~GHz maps (out of the 14 values of \thcore \
in steps of $0.25^\prime$) and should be interpreted only as a rough 
measure of that cluster's angular scale.  The value of $y_0$ reported 
for each cluster 
is the value in the 150~GHz map filtered at the best values of \thcore.
The uncertainty on the value of $y_0$ is calculated for 
\thcore \ fixed at the best value.
Once we fix $\beta$ and \thcore, $y_0$ is the only remaining 
free parameter, so the fractional uncertainty on $y_0$ is 
simply equal to the inverse of the 150~GHz S/N 
in quadrature with the 150~GHz calibration uncertainty.
}
\normalsize
\end{center}
\label{tab:detections}
\end{table*}

\subsection{X-ray Counterparts}
\label{sec:xray}

As previously mentioned, the highest significance cluster candidate in our sample
(SPT-CL 0517-5430) was identified as a galaxy cluster in
the REFLEX survey, which selected cluster candidates by correlating
sources from a ROSAT All-Sky Survey \citep[RASS,][]{truemper93}
catalog with galaxy locations from optical data.  Interestingly, 
we also find RASS counterparts to two of our other three cluster 
candidates, but these sources have not previously been considered 
as X-ray cluster candidates.  The sources are found in the 
RASS Faint Source Catalog \citep[RASS-FSC,][]{voges00}, and each lies within 1 arcmin 
of one of our SZ-detected clusters.  1RXS J054638.7-534434 lies 0.9 arcmin
northeast of SPT-CL 0547-5345, while 1RXS J050921.2-534159 lies 0.2 arcmin 
northeast of SPT-CL 0509-5342 (see Figure \ref{fig:opticalimages}).
The combined number density of RASS Bright Source Catalog 
\citep{voges99} and RASS-FSC sources in the BCS5h30 field is $6.6$
deg$^{-2}$ (roughly twice the $3.0$ deg$^{-2}$ all-sky
average), which makes the likelihood of a chance superposition within
1 arcmin of a random location in the BCS5h30 field less than
$0.6 \%$.
The RASS-FSC count rates in the broad energy band are 0.012 $\mathrm{counts}^{-1}$ for
1RXS J054638.7-534434 and 0.035 $\mathrm{counts}^{-1}$ for 1RXS J050921.2-534159.  The
count rates are both well below the flux threshold used in the REFLEX
survey, which is approximately 0.08 $\mathrm{counts}^{-1}$ in a slightly narrower
energy band.  Using the
RASS-FSC exposure times and count rates, we estimate that both of
these sources were detected with roughly 15 photons in the broad
energy band.
Neither 
of these sources would have been included in the Massive Cluster Survey 
\citep{ebeling01}, which only considered sources in the RASS-BSC.  
We find no X-ray counterpart in the literature or in publicly available 
catalogs for our cluster candidate SPT-CL 0528-5300.

\subsubsection{Positional Uncertainties in SPT SZ Cluster Detections}
\label{sec:positions}

As discussed in Section \ref{sec:maps}, the absolute pointing of the
SPT maps from which cluster candidates are extracted is accurate to
roughly 10 arcsec, as measured using sources in the PMN and SUMSS catalogs.  
However, the locations of the cluster
candidates reported in Table \ref{tab:clusterlist} will not
necessarily line up with the centers of the cluster optical or X-ray
emission to that level of precision, because the different
observations are sensitive to different combinations of physical
properties of the cluster.  
Such a separation is particularly likely in extended systems and
systems that have undergone recent mergers, such as
SPT-CL~0517-5430~/~RXCJ0516.6-5430 \citep{zhang06}.  
The position of the brightest cluster galaxy, the peak of the X-ray emission (which is
proportional to electron column density squared), and the center of
the SZ emission (which is linearly proportional to electron column
density and is smoothed by the matched filter used for
detection) can be offset by many times the positional uncertainty in 
any of the three measurements.  

\begin{figure}
\begin{center}
\includegraphics[width=0.4\textwidth]{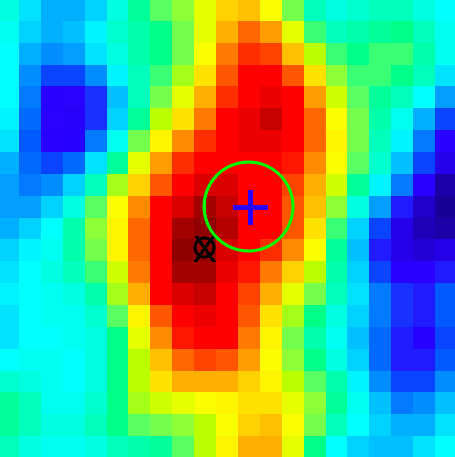}
\caption{Minimally filtered (beam-smoothed to reduce pixel noise and high-passed to remove
the primary CMB) 150~GHz image of SPT-CL~0517-5430, 
with colors and orientation as in the second row of Figure 
\ref{fig:cluster_plot}, but zoomed in such that the displayed region is 
5 arcmin on a side.
The blue cross indicates
the position of the peak of the output of the matched filter, i.e., 
the peak of the filtered 150~GHz image shown in Figure 
\ref{fig:cluster_plot}, which is also the position reported in Table 
\ref{tab:clusterlist}.  The black X and circle indicate the position of
RXCJ0516.6-5430, as reported in the REFLEX survey \citep{boehringer04}
and the positional uncertainty of the associated source in the RASS
Bright Source Catalog \citep{voges99}.  
The green 
circle around the SZ detection center is 1 arcmin in diameter as 
in Figure \ref{fig:opticalimages}, which also shows the REFLEX position.
\label{fig:sptrosat}}
\end{center}
\end{figure}

\begin{figure*}
\begin{center}
\includegraphics[width=0.72\textwidth]{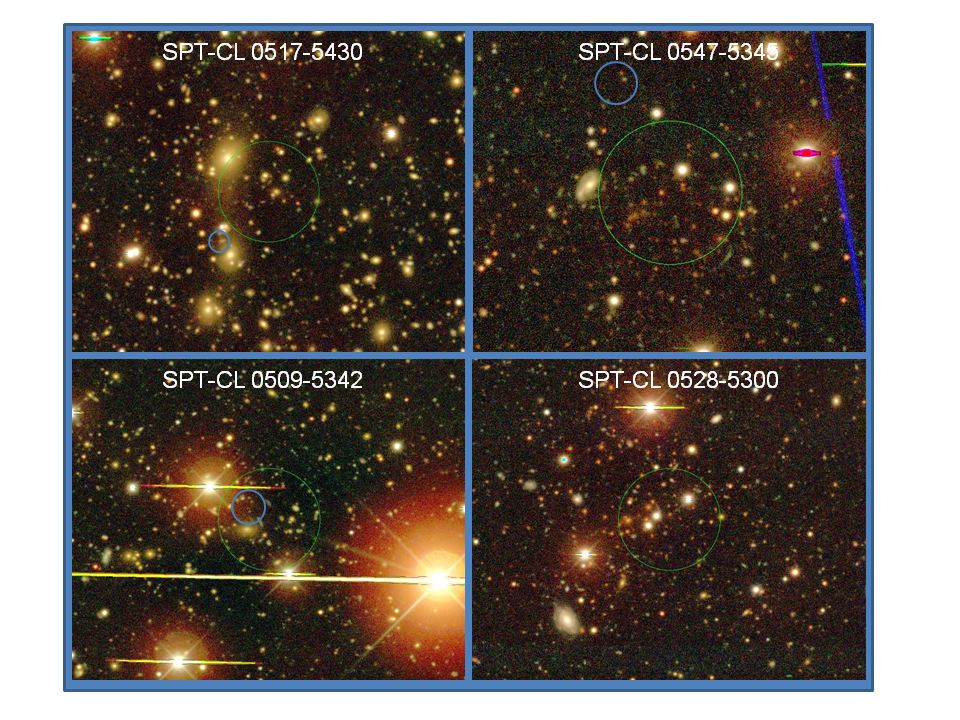}
\caption{Pseudo-color optical images of the galaxy distributions toward the SPT clusters.  All images are
oriented with north up and east to the left, as in Figures \ref{fig:cluster_plot} and \ref{fig:sptrosat}.  The SPT position is marked with a 1 arcmin diameter green circle, as in Figure \ref{fig:sptrosat}.  Populations of early-type galaxies with similar color and central giant elliptical galaxies are found to lie within 0.5$'$ of the SPT position of each system.   Gravitational lensing arcs are apparent near the central galaxy in SPT-CL 0509-5342 and to the southwest of the cluster core in SPT-CL 0547-5345.  The REFLEX position for SPT-CL 0517-5430 / RXCJ0516.6-5430 is indicated with a blue circle in the upper left panel, as are the positions of the possible RASS counterparts for SPT-CL 0547-5345 and SPT-CL 0509-5342 in their respective images.  The diameter of each blue circle is equal to the positional error given for that source in the RASS Faint Source Catalog.
\label{fig:opticalimages}}

\end{center}
\end{figure*}

Figure \ref{fig:sptrosat} shows a minimally filtered version of the 150~GHz
image in Figure \ref{fig:cluster_plot}, and it clearly shows the offset
between the center of the diffuse SZ emission and the deepest part of the
SZ decrement.  
The extent of the deepest part of the SZ decrement is 
$\sim 1^\prime$, and it is detected as a separate feature at 
$2-3 \sigma$ in this version 
of the 150~GHz map, indicating that the $\sim 0.7^\prime$ offset
between the SZ centroid and the deepest part of the decrement 
is statistically significant.
The X-ray emission should be strongly peaked in this
deepest part of the decrement, and indeed this is where the reported 
REFLEX position is located.

\subsection{Optical Counterparts}
\label{sec:optical}

We examine the galaxy distributions at the locations of the SZ selected clusters using deep, multiband optical images from the BCS.  The BCS is a 60 night NOAO survey program (2005-2008) on the Blanco 4m that has uniformly imaged 75~deg$^2$ of the sky in SDSS $griz$ bands in preparation for cluster finding with SPT and other millimeter-wave experiments.  The depths in each band are tuned to follow cluster $L_*$ galaxies to $z=1$ with $>$10$\sigma$ detections.  In addition to the large science fields, BCS covers seven small fields that overlap large spectroscopic surveys so that the photometric redshifts (photo-$z$'s) using BCS data can be trained and tested using a sample of over 5,000 galaxies.

Figure \ref{fig:opticalimages} shows pseudo-color images created from deep BCS $gri$ coadded images of the four systems.   Early-type galaxies at similar redshift appear with similar color in these pseudo-color images.  The clusters are arranged counter-clockwise by redshift starting in the upper left.  In these images the galaxy populations in SPT-CL 0517-5430 and SPT-CL 0509-5342 are yellow, and the much fainter, higher-redshift populations in SPT-CL 0528-5300 and SPT-CL 0547-5345 are orange and red.  In each image, the SPT position is marked with a 1 arcmin diameter green circle.

We have used ANNz \citep{collister04}  photo-$z$'s and color-magnitude diagrams to probe for a cluster galaxy population and to provide rough redshift estimates for each of the SPT systems.  A full analysis of the optical properties of these clusters that includes comparison to clusters selected by other means will appear in a separate paper.  Here we provide a short summary of our findings.

In SPT-CL 0517-5430, the cluster that was previously included in the Abell supplementary 
southern cluster catalog \citep{abell89} and the REFLEX catalog \citep{boehringer04}, there is a strong red sequence population apparent in $g-r$ versus $r$ and a peak in the photo-$z$ histogram.  Both indicate a redshift of $z \sim0.35$, which is roughly consistent with the value of $z=0.295$ quoted in \citet{boehringer04}.
The dominant elliptical is 0.5$'$  to the northeast of the SPT position (see Figure~\ref{fig:opticalimages}).

In SPT-CL 0509-5342 the central dominant elliptical is clearly visible $\sim0.25'$ to the southeast of the SPT position.  This galaxy is marked by two and perhaps three strong gravitational lensing arcs.  A large population of similarly colored and less luminous galaxies is apparent in the image and show up clearly as a red sequence in the $r-i$ versus $i$ color-magnitude diagram.  The red sequence indicates a cluster redshift of $z \sim0.45$, while the photo-$z$ histogram shows a peak around $z \sim0.4$, but also several peaks at higher redshift.

SPT-CL 0528-5300 has a much fainter galaxy population at higher redshift, and it appears more concentrated on the sky.  The cluster core is $\sim0.4'$ east of the SPT position, and the cluster galaxies appear orange in Figure~\ref{fig:opticalimages}.  This cluster has a red sequence visible in $r-i$ versus $i$ and in $i-z$ versus $z$ suggesting a redshift of $z \sim0.8$, again consistent with the location of a strong peak in the photo-$z$ histogram.

SPT-CL 0547-5345 is the highest redshift system; its galaxy population is faint and red in Figure~\ref{fig:opticalimages}.  The cluster core is located $\sim0.25'$ to the southeast of the SPT position.  Inspection of the images reveals a large population of faint galaxies that are not readily apparent in the figure.  The red sequence is not clearly detected for this cluster.  We note, however, that the galaxies for this cluster are close to the detection limit and therefore statistical uncertainties in their photometry are likely to be an issue.  The photo-$z$ histogram shows a peak at $z \sim0.9$, very close to the high redshift range measured for the entire field.  A feature that appears to be a strong gravitational lensing arc lies near the SPT circle to the southwest of the SPT position.
 
The BCS optical data reveal 
galaxy concentrations located within $0.5'$ of the SPT position for each of the four clusters.  We find that the cluster red sequence is present in all but the highest redshift system and gravitational lensing arcs are present in two of the SPT systems, including the highest redshift cluster.  

We also note that the trend in cluster angular size --- as evident from the images in 
Figure \ref{fig:cluster_plot} and the rough estimate of \thcore \ in Table \ref{tab:clusterlist} ---
is consistent with the redshift estimates presented in this section.  The two clusters 
estimated to lie at $z \gtrsim 0.8$ are noticeably more compact and prefer smaller values 
of \thcore \ than the clusters estimated to lie at $z \sim0.4$.  

\section{Conclusions}
\label{sec:conclusions}

In this paper, we have reported four high-significance detections 
of galaxy clusters made with the SPT, three of which are new discoveries.  
These clusters were identified first as Sunyaev-Zel'dovich effect decrements in 150~GHz SPT maps.  
The presence of decrements at the corresponding 
positions in the 95~GHz maps, and lack of signal at the same positions in the 225~GHz maps, supports their 
identification as SZ sources.  One of these four systems was previously identified 
as a cluster in the Abell supplementary southern
catalog and the REFLEX X-ray cluster catalog.  Two of the others have potential 
RASS Faint Source Catalog counterparts, but had not been identified as clusters.  
We have used data from the Blanco Cosmology Survey to produce pseudo-color optical images 
in the direction of the four SZ detections and find clear galaxy overdensities 
within 1 arcmin of the reported SPT positions.
We also see evidence for  strong gravitational lensing arcs in at least two of the optical images. 
Preliminary photometric redshift estimates 
indicate that two of the systems lie at moderate redshift 
($z \sim0.4$) and two at high redshift ($z \gtrsim 0.8$), consistent with the 
rough estimate of cluster angular scale from the SZ detections.

The cluster search presented in this paper was performed over a $\sim40$ $\mathrm{deg}^2$ subfield 
of the SPT survey region observed in both the 2007 and 2008 seasons.  
The SPT is expanding the survey coverage, with an eventual target of 
$\ge 1000 \mathrm{deg}^2$.  Future analysis will include additional data, as well 
as improvements to the data processing, calibration, and cluster identification 
algorithms.  These initial cluster detections demonstrate the potential 
of SZ effect surveys, and in particular the SPT, to produce a sample of SZ-selected 
galaxy clusters.  
In combination with optical and X-ray data, these and future SZ-selected clusters will enable 
new explorations of galaxy cluster properties and constraints on cosmological models.

\acknowledgments

The SPT team gratefully acknowledges the
contributions to the design and construction of the telescope by S.\ Busetti, E.\ Chauvin, 
T.\ Hughes, P.\ Huntley, and E.\ Nichols and his team of iron workers. We also thank the 
National Science Foundation (NSF) Office of Polar Programs, the United States Antarctic Program and the Raytheon Polar Services Company for their support of the project.  We are grateful for professional support from the staff of the South Pole station. We acknowledge S.\ Alam, W.\ Barkhouse, S.\ Bhattacharya, M.\ Brodwin, L.\ Buckley-Greer, S.\ Hansen, W.\ High, H.\ Lin, Y-T Lin, A.\ Rest, C.\ Smith and D.\ Tucker for their contribution to BCS data acquisition, and we acknowledge the DESDM team, which has developed the tools we used to process and calibrate the BCS data. We thank M.\ Brodwin, A.\ Gonzalez, J.\ Song, and C.\ Stubbs for their advice and contributions in the optical analysis . The BCS is supported by NSF awards AST 05-07688 and AST 07-08539.  DESDM development is supported by NSF awards AST 07-15036 and AST 08-13534. We thank J.\ Leong, W.\ Lu, M.\ Runyan, D.\ Schwan, M.\ Sharp, and C.\ Greer for their early contributions to the SPT project and J.\ Joseph and C.\ Vu for their contributions to the electronics.

The South Pole Telescope is supported by the National Science Foundation through grants ANT-0638937 and ANT-0130612.  Partial support is also provided by the  
NSF Physics Frontier Center grant PHY-0114422 to the Kavli Institute of Cosmological Physics at the University of Chicago, the Kavli Foundation and the Gordon and Betty Moore Foundation. 
The McGill group acknowledges funding from the National Sciences and
Engineering Research Council of Canada, the Quebec Fonds de recherche
sur la nature et les technologies, and the Canadian Institute for
Advanced Research. The following individuals acknowledge additional support:
A.\ Loehr from the Brinson Foundation; K.\ Schaffer from a KICP Fellowship; J.\ McMahon from a Fermi Fellowship;  Z.\ Staniszewski from a GAAN Fellowship; A.~T.\ Lee from the Miller Institute for Basic Research in Science,
University of California, Berkeley; and N.~W.\ Halverson from an Alfred P. Sloan Research 
Fellowship.

\bibliography{../../../BIBTEX/spt.bib}

\end{document}